\documentclass[10pt,a4paper]{article}

\usepackage{psfig}
\usepackage{latexsym}
\usepackage{amssymb}
\usepackage{epsfig,amsmath,graphics,graphicx}
\usepackage{epstopdf}

\newcommand{\LL}{\mathcal{L}}

\newcommand{\GeV}{\text{GeV}}
\newcommand{\TeV}{\text{TeV}}
\newcommand{\cm}{\text{cm}}
\newcommand{\sr}{\text{sr}}
\newcommand{\s}{\text{s}}

\newcommand{\bino}{\tilde{B}}
\newcommand{\wino}{\tilde{W}}
\newcommand{\gluino}{\tilde{g}}
\newcommand{\higgsino}{\tilde{H}}

\begin{document}
\begin{center}
{\large\bf Indirect Signals from Dark Matter in Split Supersymmetry}\\
\vskip 0.3cm {\normalsize
Asimina Arvanitaki and Peter W. Graham\\
\vskip 0.2cm
Institute for Theoretical Physics\\
Department of Physics\\
Stanford University\\
Stanford, CA 94305 USA\\
email: aarvan@stanford.edu, pwgraham@stanford.edu\\
\vskip .1in}
\end{center}

\vskip .5cm

\begin{abstract}
We study the possibilities for the indirect detection of dark matter
in Split Supersymmetry from $\gamma$-rays, positrons, and
anti-protons. The most promising signal is the $\gamma$-ray line,
which may be observable at the next generation of detectors. For
certain halo profiles and a high mass neutralino, the line can even
be visible in current experiments. The continuous $\gamma$-ray
signal may be observable, if there is a central spike in the
galactic halo density.  The signals are found to be similar to those
in MSSM models. These indirect signals complement other experiments,
being most easily observable for regions of parameter space, such as
heavy wino and higgsino dominated neutralinos, which are least
accessible with direct detection and accelerator searches.
\end{abstract}
\vskip 1.0cm

\section{Introduction}
The principle of naturalness seems to fail in explaining the small
size of the cosmological constant. The apparent fine-tuning required
dwarfs that of the gauge hierarchy problem. Further, the current
experimental bound on the Higgs mass requires a 1\% tuning in the
MSSM \cite{Dimopoulos:1981zb}, the most popular solution to the
gauge hierarchy problem. In view of these problems with naturalness,
Arkani-Hamed and Dimopoulos have proposed accepting a single
fine-tuning in the Higgs sector \cite{Arkani-Hamed:2004fb}.

In exchange, many of the problems of the MSSM, including the SUSY
flavor and CP problems and the long lifetime of the proton, are
easily solved within one framework without sacrificing either gauge
coupling unification or the natural dark matter candidate of the
MSSM.  This model, called Split Supersymmetry \cite{Giudice:2004tc},
breaks SUSY at an intermediate scale, $M_S$, between the weak and
Planck scales, giving the squarks and sleptons masses $\sim M_S$. An
approximate R-symmetry keeps the gauginos and higgsinos light,
around a TeV in mass \cite{Arkani-Hamed:2004yi}. This scale is
chosen to give a dark matter candidate whose relic abundance can be
in the observed range \cite{Lee:1977ua} and to preserve gauge
coupling unification.

Since dark matter is one of the main motivations for and constraints
on Split SUSY, it is important to understand the nature of the dark
matter candidate. We will assume that the neutralino is the LSP. The
relic abundance and direct detection signals have already been
calculated \cite{Giudice:2004tc, Pierce:2004mk}. We build on that
work to study indirect detection signals from the annihilation of
neutralinos in the dark matter halo of our galaxy.  These
annihilations produce cosmic rays which could be used not only to
discover dark matter but also to determine some of its properties.
We study the annihilation signals from $\gamma$-rays, positrons, and
antiprotons in Split SUSY. For $\gamma$-rays, we calculate the
fluxes from the center of the galaxy where the neutralino density is
greatest. Many processes contribute to the continuous flux of these
cosmic rays at tree level.  There are also one loop diagrams which
mediate the processes $\chi^0 \chi^0 \to \gamma \gamma$ and $\chi^0
\chi^0 \to \gamma Z$, creating two sharp lines in the $\gamma$-ray
spectrum \cite{Bergstrom:1997fh}. The width of these lines is set by
the kinetic energy distribution of the dark matter particles, since
photons propagate through the galaxy without significant diffusion.
The energy of these lines then serves as a very precise measure of
the mass of the dark matter particle.

We find that cosmic ray signals from neutralino annihilations in
Split SUSY are similar to those in many MSSM models.  The lines in
the $\gamma$-ray spectrum should be detectable by future
experiments, if not by current ones.  The continuous $\gamma$-ray
signal is visible for dark matter density profiles with a spike at
the galactic center.  Positrons and antiprotons from neutralino
annihilations are more challenging, as they can be observed only if
the backgrounds are well understood.  The prospects for indirect
detection are generally promising and best in areas where collider
and direct detection searches are difficult.

\section{Indirect signals}

\subsection{Split Supersymmetry framework}
To find the flux of cosmic rays from neutralino annhiliations in the
halo, we must first find the cross sections for all the neutralino
annihilation processes.  Certain differences between Split SUSY and
the MSSM are crucial to this analysis.  The most salient is the lack
of scalar superpartners at low energies. The tree level Lagrangian
in Split SUSY below the SUSY breaking scale, $M_S$, contains the
terms %%
\begin{eqnarray}
\nonumber \LL &\supset&  \bino (\kappa'_1 h^\dagger \higgsino_1 +
\kappa_2' h \higgsino_2) +\wino^a( \kappa_1 h^\dagger \tau^a
\higgsino_1 + \kappa_2 \higgsino_2 \tau^a h) - \lambda |h|^4 - \mu
\higgsino_1 \higgsino_2 - \frac{1}{2}(M_1 \bino\bino + M_2
\wino\wino + M_3 \gluino\gluino).
\end{eqnarray}
The neutralino and chargino mass matrices from this Lagrangian
are:%%
\begin{displaymath}
\mathbf{M_{\chi^0}}= \left( \begin{array} {cccc}
M_1 & 0 & -\frac{\kappa_1' \upsilon}{\sqrt {2}} & \frac{\kappa_2' \upsilon}{\sqrt {2}}\\
0 & M_2 & \frac{\kappa_2 \upsilon}{\sqrt {8}} &-\frac{\kappa_1 \upsilon}{\sqrt {8}}\\
-\frac{\kappa_2' \upsilon}{\sqrt {2}} & \frac{\kappa_2 \upsilon}{\sqrt {8}} & 0 & -\mu\\
\frac{\kappa_1' \upsilon}{\sqrt {2}} & -\frac{\kappa_1 \upsilon}{\sqrt {8}} & -\mu & 0\\
\end {array} \right),
\end{displaymath}
\begin{displaymath}
\mathbf{M_{\chi^{\pm}}}= \left( \begin{array}{cc}
M_2 & \frac{\kappa_1 \upsilon}{2}\\
\frac{\kappa_2 \upsilon}{2} & \mu
\end{array} \right),
\end{displaymath}
with $\upsilon = 246 ~\GeV$. %%
At $M_S$ the following supersymmetric relations are satisfied: %%
\begin{eqnarray}
\nonumber \kappa'_1= \sqrt{\frac{3}{10}} g_1 \sin \beta
\hspace{0.3in} \kappa'_2= \sqrt{\frac{3}{10}} g_1 \cos\beta
\hspace{0.3in} \kappa_1= \sqrt{2} g_2 \sin \beta \hspace{0.3in}
\kappa_2= \sqrt{2} g_2 \cos \beta \hspace{0.3in} \lambda =
\frac{\frac{3}{5}g_1^2+g_2^2}{8}\cos ^2 2 \beta.
\end{eqnarray}
These couplings must be run from their SUSY values at $M_S$ down to
the scale at which we are working.  This affects all the couplings
and crucially raises the Higgs mass as high as 160 to 170 GeV
\cite{Arkani-Hamed:2004fb, Arvanitaki:2004eu, Binger:2004nn}, increasing the Higgs
width to around 1 GeV.  The lack of scalar superpartners greatly
decreases the number of parameters from the MSSM.  Split SUSY models
are specified completely in terms of the input parameters $M_S$,
$\tan \beta$, $\mu$, and the gaugino masses.

\subsection{Calculation of annihilation rates}
The DarkSUSY package calculates the current relic abundance of the
lightest neutralino, the cross sections for direct detection, and
the flux of cosmic rays from neutralino annihilation in the halo
\cite{Gondolo:2004sc}.  Our calculations are based on this package
as modified in \cite{Pierce:2004mk} to include the running of the
couplings from the SUSY breaking scale, the absence of scalar
superpartners, and the effect of the process $\chi^0 \chi^0 \to h
\to WW^*$ on the relic abundance.

DarkSUSY only calculates the neutralino annihilation rate into two
body final states.  The average decay products of these two on-shell
particles are then added together to produce the indirect signal.
Normally these channels are the dominant contribution to the
annihilation rate, but since the Higgs has a large mass and width,
there is a region where the neutralinos can annihilate resonantly
through an s-channel Higgs. Even if its energy is below threshold
for WW production, a Higgs heavier than $130$ GeV has a large
branching ratio for decays to $WW^*$, which is not a two-body final
state since one of the W's is not on-shell.  When it is on
resonance, this process can dominate the neutralino annihilation
rate and significantly affect the relic abundance.

It is necessary to include the effects of this annihilation channel,
$\chi^0 \chi^0 \to h \to WW^*$, in the calculation of indirect
signals as well. However, it is suppressed by the low velocity of
the neutralinos at the present time.  In the zero velocity limit
they have no angular momentum and can only create a pseudoscalar
state in the s-channel, not a scalar.  Because of the Higgs
resonance, a velocity-suppressed signal could still be significant.
We use the known rate for the process $h \to WW^* \to Wf \overline
f$, where $f$ is any SM fermion \cite{Keung:1984hn}. The usual
DarkSUSY routines are then used to find the average decay products
of the fermions, ignoring the complication that the center of mass
frame of the fermions, the $W^*$ frame, is moving.  Taking the
motion into account would slightly affect the energies of the decay
products, but is unnecessary since we merely wish to be sure this
channel is negligible below threshold for WW production.  We find
that the velocity suppression wins and the Higgs resonance channel
contributes very little to the annihilation signals at the present
time.  In principle, the usual DarkSUSY calculation also fails for
channels with a Higgs in the two body final state, however we find
that these processes are negligible as well.

\subsection{Halo model}
The density profile of the galactic dark matter halo can affect the
calculated indirect signals greatly. We take the Burkert profile
\cite{Burkert:1995yz}:
\begin{eqnarray}
\rho (r) = \frac{\rho_0}{(1+\frac{r}{R} ) (1+\frac{r^2}{R^2} )}
\end{eqnarray}
where $\rho_0  =  0.839 \ \frac{\GeV}{\cm^3}$ and $R  =  11.7 \
\text{kpc}$. This is a conservative profile from the point of view
of annihilation signals, since it does not have a spike at the
center of the galaxy. Other halo profiles are commonly parameterized
as:
\begin{eqnarray}
\rho(r) = \frac {\rho_0} {(\frac{r}{R})^\gamma
(1+(\frac{r}{R})^\alpha)^\frac{(\beta - \gamma)}{\alpha}}.
\end{eqnarray}
These do have a spike in the center of the galaxy for positive
$\gamma$.  We will consider the Navarro-Frenk-White (NFW) model
\cite{Navarro:1996he} with $\alpha  =  1$, $\beta  =  3$, and
$\gamma  =  1$ as an alternative to the conservative Burkert
profile. We hold fixed the density at the radius of the solar system
and so take $R = 20 \ \text{kpc}$ and $\rho_0 = 0.235 \
\frac{\GeV}{\cm^3}$.  This model will tend to produce greater
annihilation signals, since they are proportional to the square of
the density. In the following sections we will compare the results
for the Burkert and NFW profiles in detail.

\begin{figure}[t]
\begin{center}
\epsfig{file=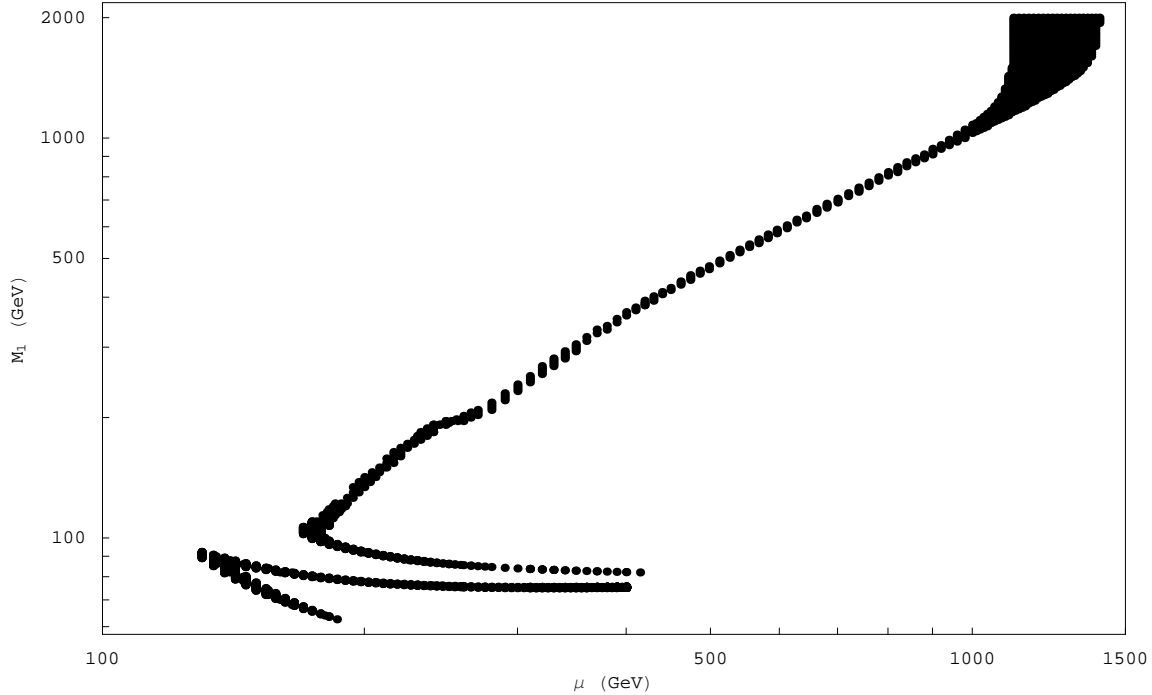, width=6.0in} \caption{ \label{Fig: Relic
Abundance} Points in the $\mu - M_1$ plane with relic abundance
within the WMAP allowed range. Here $M_S = 10^9 ~\GeV$, $\tan \beta
= 5$, and $M_2 = 2 M_1$.  The horizontal lines at low $M_1$ are the
Higgs resonance region.}
\end{center}
\end{figure}

\section{$\gamma$-ray, positron and antiproton signals}

\begin{table}[b]
\begin{center}
\begin{tabular}{|c|c|c|c|c|c|c|c|c|}
\hline
 & $M_1 (\GeV)$ & $M_2 (\GeV)$ & $\mu (\GeV)$ & $m_{\chi} (\GeV)$ & $\bino$ fraction & $\wino$ fraction & $\higgsino$ fraction & $\Omega h^2$\\
\hline
bino  & 300 & 600 & 350 & 281 & 0.730 & 0.010 & 0.260 & 0.111\\
\hline
higgsino & 1400 & 2800 & 1200 & 1192 & 0.024 & 0.002 & 0.974 & 0.109\\
\hline
wino & 4400 & 2200 & 3500 & 2197 & $\sim 10^{-7}$ & 0.997 &0.003 & 0.099\\
\hline
higgs resonance & 75.5 & 151 & 250 & 69.8 & 0.922 &  0.019 & 0.059 & 0.122\\
\hline
\end{tabular}
\caption{The characteristic bino, wino, higgsino and higgs resonance
points.  The mass, composition, and relic abundance of the LSP is
shown for each point.  For all points $\tan \beta = 5$ and $M_S =
10^9 ~\GeV$.}
\end{center}
\end{table}

In Fig. \ref{Fig: Relic Abundance} we plot the points in parameter
space that satisfy the current experimental bounds on the relic
abundance $( 0.094 < \Omega h^2 < 0.129)$ from WMAP
\cite{Bennett:2003bz}. We have chosen $\tan \beta = 5$ and $ M_S =
10^9 ~\GeV $ with $ \frac{M_2}{M_1}=2$. For low values of $M_1$,
there is a region where the neutralino annihilates through an
s-channel Higgs resonance. The neutralino in this region is almost
purely bino. Away from the Higgs resonance and as the mass of the
neutralino increases following the diagonal line, the content of the
neutralino switches from bino to higgsino.  For $m_{\chi}>1$ TeV,
the line turns upwards and the neutralino becomes almost purely
higgsino.  A wino dominated LSP can only be found at values of $M_1$
and $\mu$ larger than 3 TeV with $ \frac{M_2}{M_1}<1$, resulting in
a high mass neutralino. It should be noted that the above results
are consistent with \cite{Pierce:2004mk}.

In order to compare the continuous spectra of antiprotons,
positrons, and $\gamma$-rays in this framework with the cosmic ray
background, we consider four characteristic points in parameter
space. We calculate the fluxes for a bino, wino, and higgsino
dominated neutralino, as well as a neutralino that corresponds to
the Higgs resonance region. The masses and compositions of the LSPs
are given in Table 1.

\subsection{$\gamma$ lines}

\begin{figure}[t]
\begin{center}
\epsfig{file=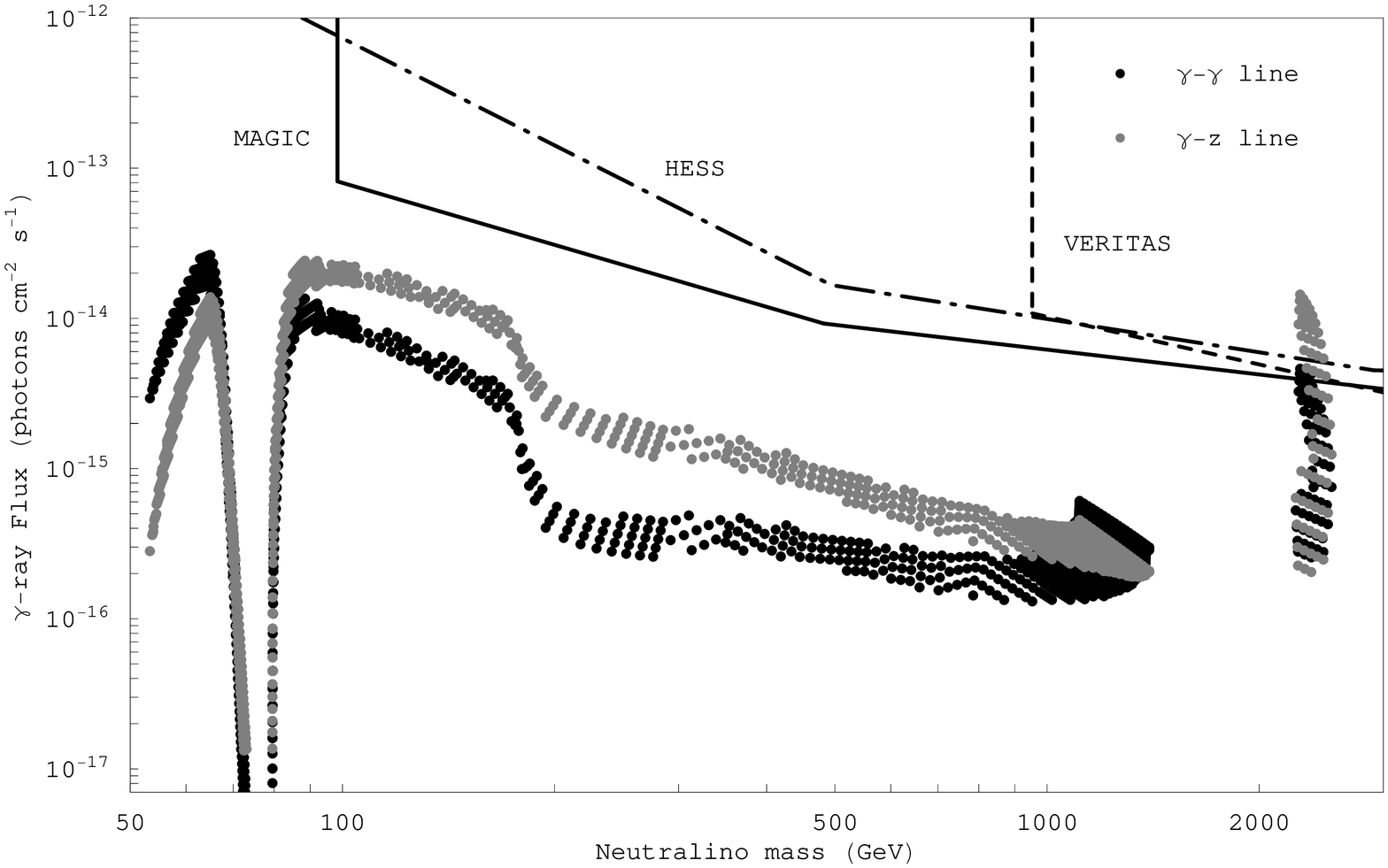, width=6.0in} \caption{ \label{Fig: Gamma
Line} The strength of the gamma lines from $\chi^0 \chi^0 \to \gamma
\gamma$ and $\chi^0 \chi^0 \to \gamma Z$ is plotted against neutralino
mass for the points in Fig. \ref{Fig: Relic Abundance} and for a scan of wino-dominated LSPs with $M_1 = 2 M_2$ (section on right).  The flux is averaged over a $10^{-5} sr$ cone around the direction of the center of the galaxy.  The sensitivities of three current experiments are shown.  We have
used the NFW profile for the signal. While the energy of the $\gamma \gamma$
line is equal to $m_\chi$, the energy of the $\gamma Z$ line is
somewhat smaller.  The suppression due to the Higgs resonance is
clearly visible.}
\end{center}
\end{figure}

\begin{figure}[t]
\begin{center}
\epsfig{file=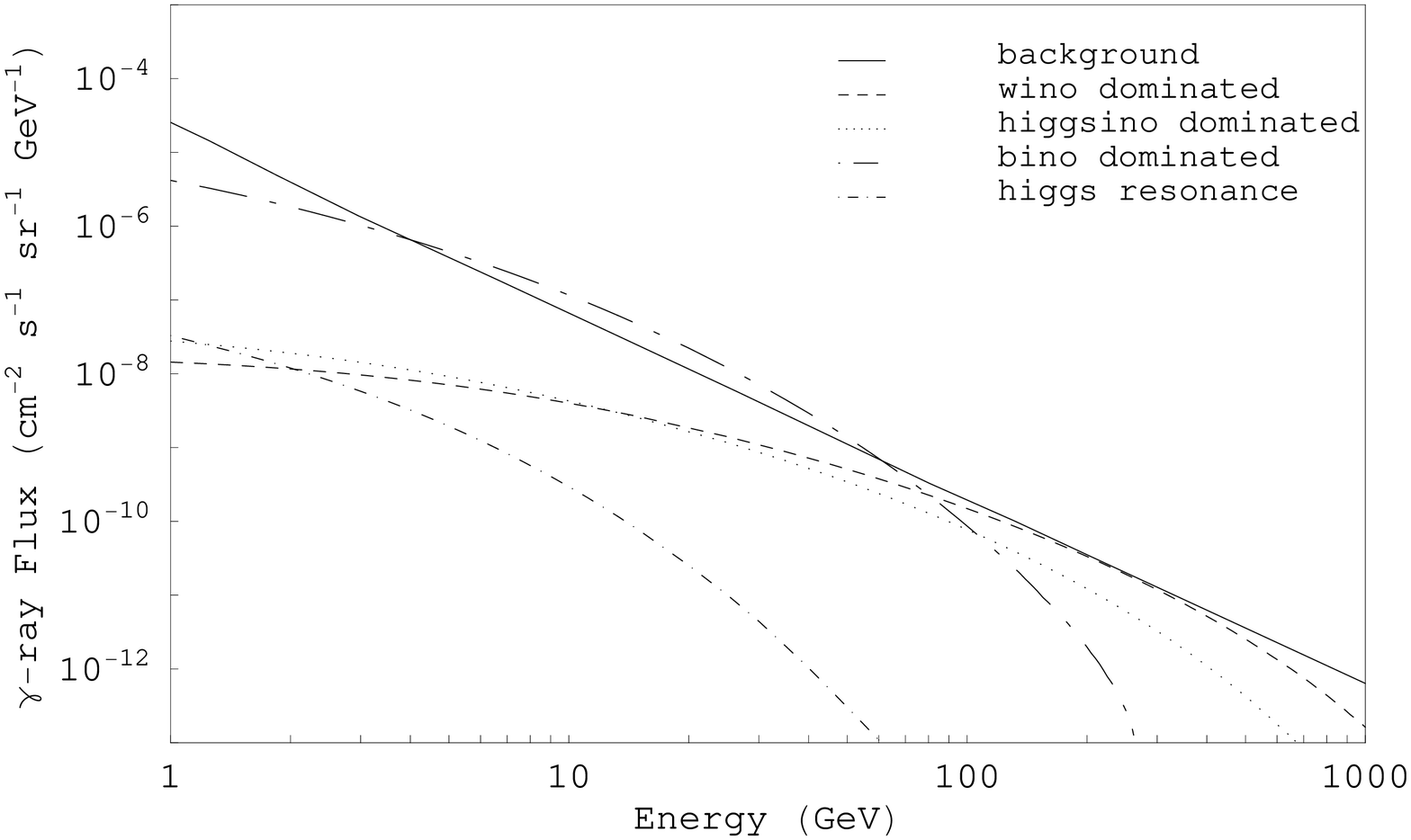, width=6.0in} \caption{ \label{Fig:
Continuous gamma ray spectrum} The continuous gamma ray spectra from
the galactic center for the four characteristic points and the
expected background.  We have used the less conservative NFW profile
for the signals.}
\end{center}
\end{figure}

The annihilations of two neutralinos to two $\gamma$-rays or to a
$\gamma$ and a Z produce lines in the $\gamma$-ray spectrum which
can provide a striking signature of the dark matter particle. The
energy of the $\gamma\gamma$ line is $E_{\gamma}=m_{\chi}$, since
the neutralinos have low velocities. The $\gamma Z$ line occurs at
an energy $E_{\gamma}=m_\chi-\frac{m_Z^2}{4m_{\chi}}$. This is
practically indistinguishable from the $\gamma \gamma$ line for
large neutralino masses, where the two will be easily within any
current experiment's energy bin size.

Unlike positrons and antiprotons, $\gamma$-rays can travel through
the halo with little diffusion, especially if they are highly
energetic. This makes probing distant sources that have
increased local density, such as the center of the galaxy, possible.
Now, the halo profile chosen for our calculations starts to play an
important role. As already discussed, many halo models favor the
existence of a spike in the dark matter profile at the galactic
center. We calculate the flux from this direction in the sky averaged
over a solid angle of $10^{-5} \ \sr$, which we assume to be the
angular resolution of our detector.

Fig. \ref{Fig: Gamma Line} shows the expected flux of the two
$\gamma$ lines from the direction of the galactic center as a
function of the neutralino mass, for all the points shown in Fig.
\ref{Fig: Relic Abundance}.  It also includes a scan of wino-dominated neutralinos (the strip on the right) with the same parameters except $M_1 = 2 M_2$.  Here we have used the NFW profile.  For the Burkert profile the signals are reduced by a factor of about 1000. The Higgs resonance causes a significant
suppression of the signal at $m_\chi \approx \frac{1}{2} m_H \approx
80 ~\GeV$.  The $\gamma$ line signals are peaked at $\approx
3 \times 10^{-14} ~\cm^{-2} \s^{-1}$. This signal is not observable
by space-based experiments, since even GLAST, with an effective area
of $\sim 10^4 \ \cm^2$ \cite{Morselli:2003xw}, will see only $\sim
0.01 ~\text{photons} ~\text{yr}^{-1}$ from the galactic center and $\sim 1 ~\text{photons} ~\text{yr}^{-1}$ from the entire galaxy. Ground based telescopes, such
as HESS \cite{Aharonian:1997}, have a much larger effective area
($\sim 10^8 ~\cm^2$), allowing a possible detection of the signal.

Of course, the astrophysical background could also overwhelm the
signal, though a line will be much easier to detect than the
continuous spectrum.  Fig. \ref{Fig: Gamma Line} includes the sensitivities of three currently operating telescopes.  These sensitivities are calculated for a 1 year exposure from the known integral point source sensitivities \cite{Morselli:2003xw, sensitivities}.  We expect the actual sensitivity to be better because our signal is not just a point source but also a monochromatic line which aids in distinguishing it from the background.  However, the quoted sensitivity should be approximately correct because the background falls sharply with energy.  If a line source is observable above the background at its own energy, it is probably observable above the total integrated background greater than its energy.  Further, a calculation of the sensitivity to a line source \cite{Bergstrom:1997fj} gives a sensitivity for HESS and VERITAS which is very similar to the quoted sensitivities.  From Fig. \ref{Fig: Gamma Line} it is clear that MAGIC has the best sensitivity in the relevant energy range although HESS is quite similar and VERITAS is just as sensitive above $1 ~\TeV$.  HESS and VERITAS are similar air-Cherenkov telescopes, but VERITAS is in the northern hemisphere.  Since the galactic center is a southern source, VERITAS observes it at a high zenith angle giving a higher energy threshold, but better sensitivity.

For high mass neutralinos the signal remains roughly constant, in contrast to the background which falls as a power law as in Fig. \ref{Fig: Continuous gamma ray spectrum}.
Unfortunately, the detectors have large energy resolutions, around
10-25\%, which must be factored in when comparing the signal in Fig.
\ref{Fig: Gamma Line} to the background, making it harder to see a line.  For the NFW profile many of the heavier wino-dominated points should be visible at current experiments, especially since the $\gamma-\gamma$ and $\gamma$-Z signals should be added together as they will not be distinguishable in current telescopes.  The rest of the points are almost visible and will be probed by next generation experiments.  In addition, since there are multiple experiments operating in this energy range, a discovery could be confirmed easily.  Of course, these results depend strongly on the halo profile and so the improvements in sensitivity of future telescopes will really be allowing us to probe halo profiles with less dark matter at the center of the galaxy.

Using the Burkert profile reduces the signal from the galactic center by a factor of 1000 making it probably undetectable at current or next generation telescopes.  In this case it would actually be much easier to observe the signal by observing the entire galaxy instead of just the center because the density is roughly constant over the galaxy.  Thus telescopes like GLAST with a large field of view have an advantage over telescopes which must be focused on a small portion of the sky.  Since GLAST will see at most $\sim 1 ~\text{photon} ~\text{yr}^{-1}$ in the $\gamma$-line for either the Burkert or NFW profiles, it is probably not enough for a detection.  But because this is about the expected level of the background for neutralinos with mass $\gtrsim 1 ~\TeV$, a similar next generation telescope might even be able to detect the line signal in the case of a halo profile without a spike at the galactic center.

The calculated $\gamma$ line signals are similar in strength to
those of MSSM models, being slightly smaller than the signal from
Anomaly Mediated SUSY Breaking (AMSB) models but larger than that
from mSUGRA models \cite{Hooper:2003ka}.  In AMSB models the LSP is
either wino or higgsino dominated and so the comparison should be
with those points.  The mSUGRA models can also have a bino LSP and
allow a general comparison with Split SUSY. The new feature in Split
SUSY is the Higgs resonance region where the flux is greatly
suppressed. Away from this region, the prospects for the detection
of the $\gamma$ line in Split SUSY are much the same as in the MSSM.

\subsection{Continuous $\gamma$-ray, antiproton and positron flux}

\begin{figure}[t]
\begin{center}
\epsfig{file=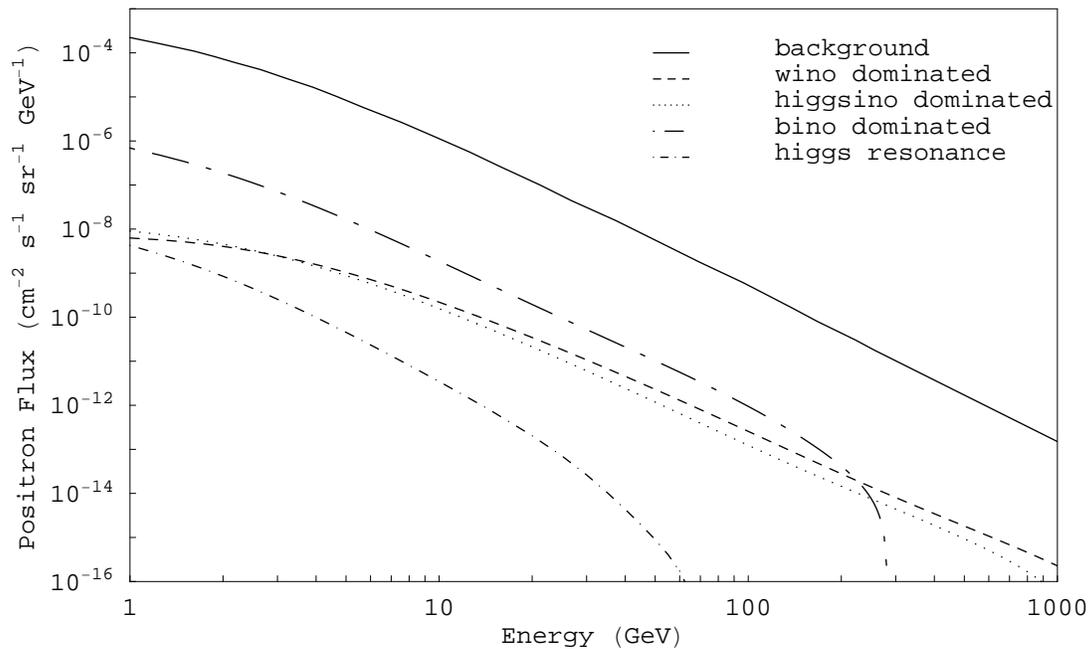, width=6.0in} \caption{ \label{Fig:
Continuous positron spectrum} The continuous positron spectra for
the four characteristic points and the expected background.  We have
used the NFW profile though it will not affect the signals
greatly, since they depend primarily on the local halo density.}
\end{center}
\end{figure}

\begin{figure}[t]
\begin{center}
\epsfig{file=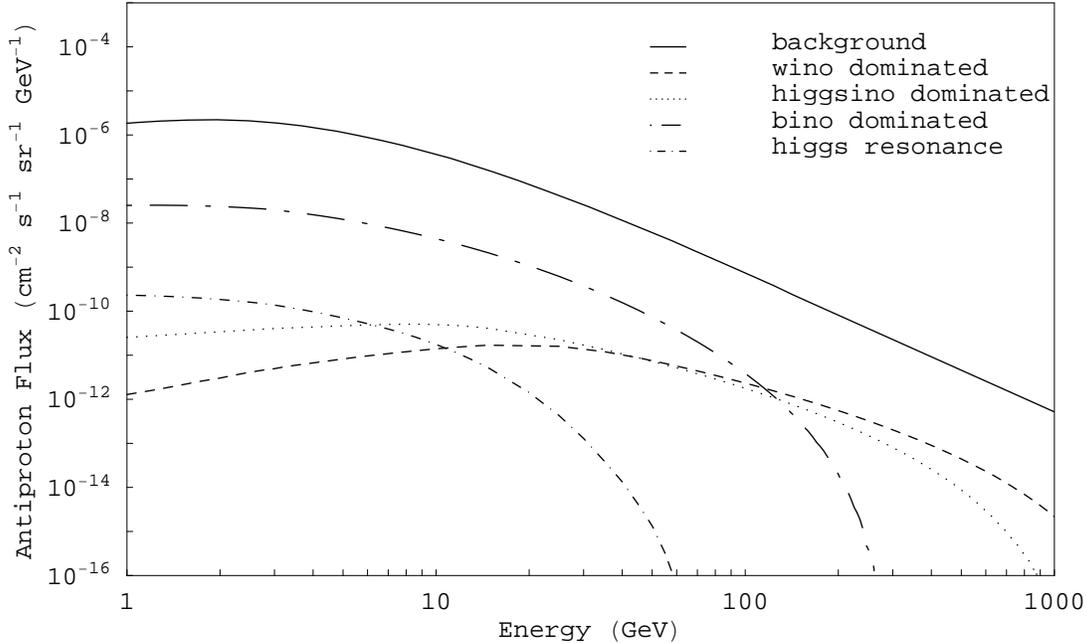, width=6.0in} \caption{ \label{Fig:
Continuous antiproton spectrum} The continuous antiproton spectra
for the four characteristic points and the expected background.  We
have used the NFW profile though it will not affect the signals
greatly, since they depend primarily on the local halo density.}
\end{center}
\end{figure}

The continuous spectrum of $\gamma$-rays, antiprotons and positrons is mainly produced from the
decay of annihilation products and, in particular for $\gamma$-rays, the radiative emission of photons
when those annihilation products propagate through the halo. While $\gamma$-rays make it possible to probe the galactic center directly, charged antimatter can only propagate a few kpc, and the signal will be determined by the known local halo density.

In Figs. \ref{Fig: Continuous gamma ray spectrum},  \ref{Fig: Continuous positron spectrum}, and \ref{Fig:
Continuous antiproton spectrum}, we present the differential yields for the four characteristic points, as evaluated using the NFW profile. We have used the solar modulation routines in DarkSUSY to account for the propagation of antimatter in our solar system. In the Higgs resonance
region the signal is strongly suppressed, as expected, because the LSP is almost purely bino and has reduced coupling strength. The signals from all other points are comparable in comparison to the
background, merely shifted towards higher energies for the
higher mass LSPs. The flux goes to zero as the energy approaches
$m_{\chi}$. Even though the number of tree level diagrams  contributing to the yield is reduced by the absence of the scalar superpartners, the calculated signal is not much smaller than in
ordinary SUSY models \cite{Ullio:2001qk, Profumo:2004ty,
Feng:2000zu}. LSP annihilations in an AMSB model provide a slightly
larger signal.

The calculated signal for the $\gamma$-ray spectrum with the NFW profile is comparable to the background while the antimatter signals are at least two orders of magnitude smaller than their backgrounds \cite{Morselli:2003xw,Boezio:1997ec,Orito:1999re,DuVernois:2001bb}.  For $\gamma$-rays, this signal should be observable by both ground and space based telescopes. The signal is suppressed without a spike in the density at the galactic center. For the Burkert profile, this suppression translates to a factor of $\sim 10^{-3}$. This signal is below the statistical errors for the background and so unobservable for experiments like GLAST and at experimental sensitivity for ground based telescopes like VERITAS. For antiprotons and positrons, the signal cannot be conclusively observed by current or next generation experiments.  Since charged antimatter particles cannot reach us from the center of the galaxy, the calculated fluxes will not be affected by the presence or absence of a cusp.

To determine whether these signals are observable, it is also important
to understand the background model. There is a lot of
uncertainty about the interstellar medium in which the diffuse
$\gamma$-ray background and antimatter particles propagate.  There have been efforts to create a viable model that agrees with the latest data \cite{Strong:2004de}. More data from
current and next generation experiments can improve our
understanding of the background spectrum.  In the absence of a well
understood background model, it may be impossible to see a
continuous $\gamma$-ray or antimatter signal that is a couple orders of magnitude
below background, even if it is above the statistical error.

Although the spectrum produced is similar in shape to the background
there are several features which should allow a detection.  The
annihilation signal has a characteristic rise compared to the
background and then decreases abruptly at the neutralino mass.  If
there is a central spike in the density profile, the $\gamma$-ray
signal should be readily observable.  With a good enough
understanding of the background, even the antimatter or the $\gamma$-ray signal from the Burkert
profile could be observable.

\section{Conclusion}

We have discussed the basic aspects of indirect dark matter signals
in the Split SUSY scenario.  In general, the situation is similar to
that in the MSSM, though with more restrictions on the possible
models. The positron and antiproton spectra probably cannot be
detected due to the large uncertainty in the background and the low
annihilation signal produced.  The continuous $\gamma$-ray signal
will be difficult to observe for conservative halo profiles.
However, a central spike in the galactic halo density could make
this signal observable at current or near-future detectors.

The $\gamma$-ray lines are easier to detect than the continuous
spectra, since there are no other known sources for such a signal.
Current experiments can already probe certain regions of Split SUSY
parameter space with a less conservative halo profile.  For halo profiles which are sharply peaked, MAGIC and HESS are sensitive to neutralino masses above $\sim 100 ~\GeV$ and VERITAS to masses above $\sim 1 ~\TeV$.  Since the diffuse $\gamma$-ray background falls with energy, experimental sensitivities improve at high energies, making the high mass higgsino- and wino-dominated neutralinos more easily observable.  Future
experiments should be able to see the $\gamma$-line signal for a wide range of
the available parameter space and halo profiles.  This could provide
not only a discovery of the dark matter particle, but also a
measurement of its mass.

Decays of gravitinos can produce an additional, non-thermal
abundance of LSPs in Split SUSY \cite{Arkani-Hamed:2004yi}, which we
have not included in our calculations.  We expect this to increase
the relic abundance and thus change slightly the allowed regions of
parameter space from that shown in Fig. \ref{Fig: Relic Abundance}.
In general, this will increase the number density and decrease the
mass of the LSP, leading to enhanced indirect signals compared to
those presented above.

These indirect dark matter searches tend to complement accelerator
and direct detection experiments.  Of course, indirect signals can at best provide a detection of the dark matter particle and a measurement of its mass.  In order to discover supersymmetry or distinguish Split Supersymmetry from other supersymmetric models, further evidence will be required for example from colliders \cite{colliders}, cosmic ray observatories \cite{Anchordoqui:2004bd}, or neutrino telescopes.  A lower mass LSP will be
discovered at the LHC, while a heavier one gives the best
annihilation signals.  Pure higgsino or wino dark matter would be
very difficult for direct detection and collider searches
\cite{Pierce:2004mk}, but should be detectable indirectly. In this
case, astrophysics can fill in the pieces of the puzzle that other
experiments have missed.

\section*{Acknowledgements}
We would like to thank T. Abel, E.A. Baltz, P. Michelson, and S.
Thomas for useful discussions.  Special thanks to C. Davis, S.
Dimopoulos, A. Pierce, and J.G. Wacker for all their help.  P.W.G.
is supported by the National Defense Science and Engineering
Graduate Fellowship.

\end{document}